# Multiscale modelling of tumour growth induced by circadian rhythm disruption in epithelial tissue [1]

D. A. Bratsun  D. V. Merkuriev
A. P. Zakharov  L. M. Pismen

**Abstract** We propose a multiscale chemo-mechanical model of cancer tumour development in an epithelial tissue. The model is based on transformation of normal cells into the cancerous state triggered by a local failure of spatial synchronisation of the circadian rhythm. The model includes mechanical interactions and chemical signal exchange between neighbouring cells, as well as division of cells and intercalation, and allows for modification of the respective parameters following transformation into the cancerous state. The numerical simulations reproduce different dephasing patterns – spiral waves and quasistationary clustering, with the latter being conducive to cancer formation. Modification of mechanical properties reproduces distinct behaviour of invasive and localised carcinoma.

---

[1] The research has been supported by the Ministry of Education of Perm Region (grant C-26/244) and the Russian Fund for Basic Research (grant 14-01-96022r_ural_a).

A. P. Zakharov    L. M. Pismen
Department of Chemical Engineering, Technion - Israel Institute of Technology,
32000, Haifa, Israel
Tel.: +972-4-8295671
Fax: +972-4-82930860
E-mail: pismen@technion.ac.il

D. A. Bratsun
Theoretical Physics Department, Perm State Humanitarian Pedagogical University,
614990, Perm, Russia

D. V. Merkuriev
Department of Hospital Pediatrics, Perm State Medical Academy,
614990, Perm, Russia

# 1 Introduction

Cancer modelling has, over past decades, become really one of the challenging problems for mathematicians, physicists and other researchers working in the biological sciences [1].

One of the main problems of mathematical modelling of cancer (as, indeed, of any biological system) is the multiscale nature of this phenomenon [2, 3, 4, 5]. One can identify at least three characteristic spatial scales including the microscopic (the size of cell's nucleus, less than 10 m $\mu$), mesoscopic (cell's size, 10-100 m $\mu$) and macroscopic scale (the size of the tumour or the organism as a whole). At the microscopic scale, a system of ordinary differential equations can be used to simulate the system within the population dynamics, where each variable corresponds to a some biological property characteristic to all cells of the same population. The paper [6] seems to be the first to apply of population dynamics models focused on cancer. This direction has been further developed to account for more fine effects (see, for example, the survey [7]). The advantage of this approach is that the models are easily manipulable, and enable a relatively fast determination of the governing parameters. It omits, however, spatial effects and some other important aspects. Another type of modelling, supplementing the population dynamics by certain variables determining the average structure of a population of cancer cells (for example, the age of cells), can be attributed to the class of semi-phenomenological models with internal structure [8].

A large body of literature has been dedicated to models linking the microscopic scale to the tumour scale [2]. This approach allows one to understand how changes in cell-cell communications influence macroscopic properties of the malignant growth. Most commonly, the tumour is considered as a continuous medium. It is described by a system of partial differential equations including the mass balance equation for the cellular medium and reaction-diffusion equations describing the field of signalling between cells. The tumour may be viewed as a solid porous matrix [9] which interacts with a filling cellular liquid of healthy cells. In this way, one can describe some spatial features of a real tumour (for example, its fractal structure). In a more recent work of this kind [10], the tumour was represented as a multiphase medium simulated in a three-dimensional geometry by a finite elements method. A disadvantage of such models is their inherently phenomenological character.

An alternative approach is cell-based discrete modelling. In contrast with continuous medium models, discrete models enable to trace the functioning of individual cells. Thanks to the progress in biotechnologies, there is an increasing amount of experimental data available at a single cell level which should be taken into account in mathematical models. Simpler models of this kind are based on cellular automata [11] or random walk of cells [12]. The latter approach allows one to consider stochastic processes of tumour formation but cannot account for their mechanical properties. A large group of models combine interactions at both sub-cellular and cellular level [2, 4]. This approach is motivated by the fact that the biological system as a whole is driven by genetic mutations: for example, too strong expression of certain genes can lead to malfunction of the cells, i.e. transition to the cancerous state [13]. Stochastic description of gene regulation processes can be naturally applied here [14].

Since cancer is a multiscale phenomenon, the most realistic approach to modelling requires involving processes at all levels of description. Due to the difficulties of this approach, there are not so many works of this kind. One of the first attempts to implement a hybrid approach included a model of cellular automata whose state is determined by a continuous distribution of oxygen around a blood vessel near the origin of the tumour [15]. In another remarkable work [16], the authors examined a spherically growing tumour within the framework of a lattice model of mechanically interacting discrete cells, in which the gene regulation processes have been simulated separately for each cell in the population. A good review of recent developments in multiscale cancer modelling can be found in Ref. [4]. We would also like to mention the book [5] covering state-of-the-art methods of multiscale cancer modelling.

We conclude that any realistic model of tumour growth should involve a dynamical model of interaction of a large number of cells [17]. Such a model must take into account the physical properties of individual cells in the ensemble, *e.g.* to describe dynamically changing volumes and surfaces of cells, their elastic response to external mechanical impacts, the ability to move, divide, etc. These processes must be influenced by the exchange between the cells of the various signals generated at either subcellular (transcription/translations) or macroscopic level (collective behaviour). The model should combine therefore a discrete description of individual cell dynamics and a continuous description of chemical fields that are common to the whole ensemble. The development of such models is not easy but the progress in this area over the last decade and the simultaneous breakthroughs in the computer technology are bringing the researchers closer to the realistic simulation of the living tissue functioning.

This communication combines modelling of signalling regulation of cancer formation with cell-based mechanical modelling of deformations and rearrangement of epithelial tissue. It has been recognised in recent years that core circadian genes are important in tissue homeostasis and tumourigenesis. In recent years, a significant number of works have shown that disruption of the circadian mechanism is implicated in gene deregulation leading to the development of cancer and other diseases [18, 19, 20]. The main idea of this work is that the tumour occurance is induced by a local disruption of the circadian rhythm in the epithelial tissue when the control of the peripheral circadian oscillators by the suprachiasmatic nuclei weakens or vanishes completely. We have supplemented the mechanical epithelium model which was developed earlier [21] by two simple models of circadian rhythms suggested in our previous papers [22, 23, 24], where the crucial element of the mechanism of oscillations is a delay of protein synthesis during the process of gene regulation. These constitutive parts are supplemented by a simple phenomenological model of the cell transformation into the cancerous state leading to modification of its physical properties.

# 2  Description of the model

## 2.1  General principles

Epithelium can be defined as a relatively avascular aggregation of cells which are in apposition over a large part of their surfaces, and are specialised for absorptive, secretory, protective, or sensory activities [25]. Epithelia may occur as sheets of cells, characterised as a covering and lining epithelium, as in the lining of the intestine, or as solid aggregations of cells, as in glandular organs. We focus our attention on the first type, and construct a two-dimensional model of a single layer of epithelial cells. Epithelial cells can be shaped in diverse ways. The three basic cell shapes in covering and lining epithelia, distinguished by their appearance in microscopic sections, are squamous, cuboidal, and columnar but the distinctions are often blurred. The cells adhere to each other by forming specialised attachment structures (adherens junctions or desmosomes) that ensure coherence and strength of tissue.

We will further concentrate on modelling of carcinomas. Carcinoma is a type of cancer that develops from epithelial cells. It may affect the skin, colon, prostate, breast, and lung, and is among the most widespread forms of cancer in adults, caused when DNA is altered or damaged to such an extent that the cells start to exhibit abnormal malignant properties. The following properties of cancer cells will be taken into account in our model:

- a normal cell can turn into a cancer cell ([1], p.7);
- cancer cells do not undergo reverse transformation ([1]; p.7);
- cancer cells exhibit a number of alterations of their physical properties in comparison to normal cells ([1], p.285);
- cancer cells show uncontrolled mitotic divisions causing disorganised growth ([1], p.45; [26, 27]);
- cancer cells are far less adhesive than the normal cells, and therefore tend to wander through tissues causing cancerous growth in different parts of the body ([1], p.215);
- cancer cells exhibit a number of alterations in their genes: for example, their circadian rhythm is disrupted, and they cease to exchange signals with normal cells ([1], p.39; [26, 27, 28]).

In order to arrive at a realistic description enabling to obtain stress field and velocity distribution in the epithelium, we have developed a discrete chemo-mechanical model. The main components of our model are *mechanical*, involving elastic interactions between cells, their proliferation and division, and *genetic*, involving the mechanism of circadian rhythms and intercellular signalling. The mechanism of the cell transformation is the third important element of the model. Their relationships are schematically shown in Fig.1.

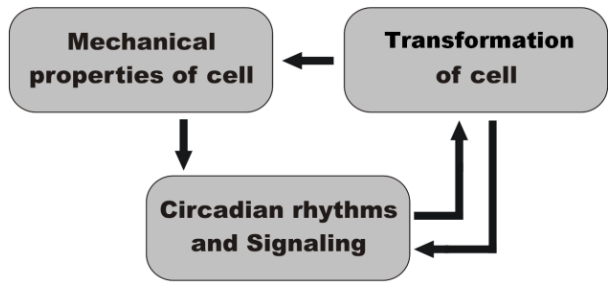

Figure 1: Principal components of the model and their relationships

The mechanical submodel presents the epithelium as an elastic two-dimensional array of cells, approximated by polygons. The evolution of the system starts from a hexagonal cells constituting the plane. In the course of proliferation, division and transformation, the initial lattice deforms incorporating also irregular polygons. The deformation of the epithelial layers composed of cells that tightly adhere to one another, is a crucial process in tissues since it plays a central role in gastrulation, early phase in the embryonic development that initiates the formation of three-dimensional structures of tissues and organs [29]. The physical properties of cancer cells significantly differ from those of the normal cells [1, 26, 27]. Thus, the transformation of the cell into cancerous state changes the mechanics of cell behavior. This assumption establishes the important link between the elements of the model shown in Fig.1.

The key feature of our genetic model is the influence of periodic inputs. *Biological rhythms* are periodical fluctuations, which are quite characteristic for all forms of life on earth. Rhythms are closely connected adaptive phenomena to periodic changes in environmental factors. We concentrate here on *circadian* rhythms that synchronise with daily changes of the environment. A remarkable feature of these rhythms is that they are produced by autonomous intracellular mechanism [30].

To date, it is widely accepted that the genetic mechanism is responsible for circadian oscillations [31]. It is now realised that circadian rhythms manifest themselves even on the subcellular scale in the form of RNA and protein fluctuations. As soon as transport proteins pass through the cell membranes and trigger intracellular interactions, circadian oscillations inevitably develop at the mesoscopic and macroscopic scales. At the organism scale, the signals from individual cells are synchronised, thus forming generalized rhythm for the organism as a whole. Now it is established that the main circadian rhythm in mammals is produced by the suprachiasmatic nuclei (SCN), the group of cells located in the hypothalamus. One of the principal discoveries was the finding of ultra-fast synchronization of oscillations within 2-3 periods due to strong coupling between SCN neurons via a unique mechanism of neuropeptidergic signalling [32]. As a consequence, any pattern formation cannot occur in such a medium: SCN maintain full synchrony as a population [33]. It may imply that the circadian clock of higher organisms can behave differently [34, 35] . For example, the authors of [35] have proposed to use phase response curve analysis to show that the coupled period within the SCN stays near the population mean instead of a Hill-type regulation widely assumed in previous mathematical models.

Though the SCN clock is the circadian pacemaker for the whole organism, the peripheral circadian oscillators in mammals are not just a passive medium. It was shown that peripheral clocks are cell-autonomous and can produce rhythm irrespective of SCN [36]. Thus, the master clock rather synchronizes peripheral clocks which work locally and autonomously, than sends the direct commands to control the local circadian physiology.

The problem of the spatial synchronisation of a large number of interacting oscillators is actively discussed in the physical literature [37]. Biological applications were hindered, until recently, by the lack of experimental data but the situation has begun to change due to the appearance of sophisticated experimental techniques. Some examples are gene expression data analysed from postmortem brains [38], and attempts at spatial analysis of fluorescent genes [39]. But still most studies largely concentrate on the temporal rather than the spatial organisation of rhythms.

The principal issue for pattern formation is a local coupling between clocks in the peripheral cells. Although peptidergic signalling is absent in the peripheral tissues, other mechanisms should be present. For example, it was reported in [40] that the peripheral tissues of SCN-lesioned mouse still demonstrate some degree of synchronization in the local circadian clocks. Of course, the coupling between peripheral oscillators is much more weaker that in SCN, but it still exists [36]. It means that when the SCN control disappears, the weak coupling of local oscillators leads to greater freedom of each of them in a whole group, and hence to the possible formation of collective patterns.

Since there is extensive experimental evidence that gene deregulation influenced by circadian clock is involved in the development of cancer [18, 19, 20], the molecular mechanism of the circadian rhythm disruption and its propagation due to cell-to-cell signalling should become an essential factor in transformation of cells. In mammals, it is known now that if the connection between the peripheral clocks and the pacemaker in SCN is destroyed the organism becomes at high risk of cancer. For example, it was demonstrated in [41] that when the peripheral tissues in mice deduced from the SCN control were artificially synchronized by meal timing, the cancer growth was reduced by 40%.

Analyzing all the above, one can conclude that without the SCN control the local circadian oscillators are left to themselves, and can weakly interact with each other, forming poorly synchronized patterns which induce locally the cell transformation. This assumption establishes the principal link between these two elements of the model (Fig.1). This link operates in both directions: on the one hand, the circadian patterns cause the cell transformation ([28, 42, 43], see discussion below); on the other hand, the circadian rhythms in cancer cells are either switched off or seriously damaged ([26, 27, 28]).

## 2.2 Mechanical cellular model of epithelial tissue

We adopt a two-dimensional mechanical model of the epithelial layer along the lines of an earlier study of wound healing [21]. The cells always remain attached to each other forming a continuous two-dimensional epithelial surface (Fig. 2). The curvature of the layer (presumed small compared to the cell size) and thickness inhomogeneities are neglected. The following features of the model described in detail below make it suitable for realistic simulations of the epithelium:
- allowing for changing the size and shape of cells;
- allowing for tissue spreading via the mechanism of cell division (Fig. 2a);
- allowing for motion of cells by the mechanism of intercalation (Fig. 2b);
- including the dynamics of signalling species taking part in the regulation of intracellular processes;
- including the exchange of chemical signals between adjacent cells through their shared borders (Fig. 2c).

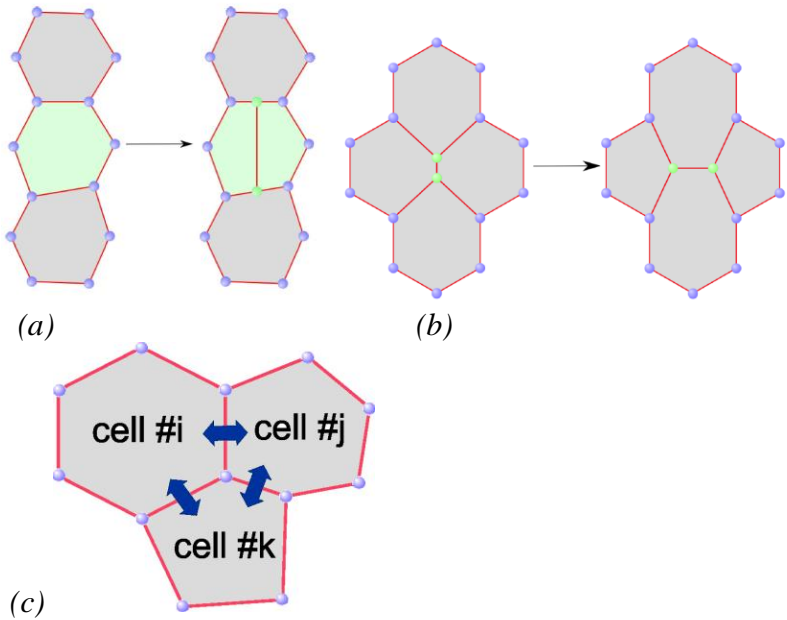

Figure 2: Elements of chemomechanical model of epithelium: *(a)* - cell division; *(b)* - cell intercalation; *(c)* - cell-to-cell communications

The mechanical model is based on the elastic potential energy $U$ of the tissue [44]:

$$U = \frac{1}{2}\sum_{cells}\left(\mu P^2 + \eta(A - A_0)^2\right), \tag{1}$$

where $P$ and $A$ stand for the perimeter and area of cell respectively. Here the coefficient $\mu$ characterises the effect of contractile forces within the perimeter of the cell, $\eta$ characterises the elastic resistance to stretching or compressing the cell with respect to the reference cell area $A_0$.

Vertices of the polygons representing the cells form a lattice. Evolution of the tissue is modeled by moving the lattice nodes (Fig. 2). We define the mechanical force acting on any $j$ th node in a standard way:

$$F_j = -\frac{\partial U}{\partial R_j}, \qquad (2)$$

where $R_j$ denote the position of the node.

We consider the internal movement of cells in the tissue as strongly overdamped process, so the equation of motion for $i$ th node can be written as

$$V_i = \frac{dR_i}{dt} = KF_i H(|F_i| - F_0), \qquad (3)$$

where $V_i$ is the velocity, $K$ is the mobility coefficient, $H$ is the Heaviside function. The threshold force $F_0$ has been introduced in (3) to take into account the situation when the node remains immobile even if force is not zero.

An important feature of the model is the ability of cells to divide. This allows the tissue to carry out internal movement through the redistribution of internal stresses in the environment. It is assumed that the division occurs when the longest edge of the polygon and its corresponding opposite edge are divided in half (Fig. 2a). In order to minimize the growing disorder in the distribution of nodes in the process of cell division, the probability to divide was connected to the number of vertices of the polygon $n_j$ according to the following formula (see [21] for more detail):

$$p_{div} = p_0 q^{n-6}, \qquad (4)$$

where $q$ is a distribution constant and $p_0$ is a scaling factor. One can see that for $q > 1$, the polygons with a number of sides over 6 experience the division more frequently. Thus, the most likely shape of the cell according to (4) is a hexagon. For cancer cells, the dependence of the number of nodes is suspended and the division rate is set at considerably higher level.

In fact, the real epithelial tissues can demonstrate the effect of extrusion of some cells leading to their death. There is also increasing evidence that clones of cells exhibiting differential proliferation rates, such as those induced through oncogenic transformation, lead to cell competition effects [45]. For sake of simplicity, we here neglect all these phenomena.

Another mechanism which increases the liquidity of the tissue and excludes the severe deformation of some cells is intercalation [21, 46]. We introduce here a special parameter $l_0$ which determines the moment when the intercalation can occur. Then the probability of the event can be written in the simple form:

$$p_{int} = \begin{cases} 1, & l_i < l_0 \\ 0, & l_i \geq l_0 \end{cases}. \qquad (5)$$

One can see from (5) that if the length of the border separating two cells becomes less than $l_0$, it is substituted by a link of a slightly larger length in the normal direction (Fig. 2b). The intercalation is known to be important in many tissue reshaping processes. Since cancer cells are far less adhesive than normal cells, the respective $l_0$ is set at a higher level, thereby making the intercalation of cancer cells more probable. Altogether, Eqs. (1-5) define the mechanical dynamics of tissue on the cellular level.

## 2.3 Genetic models of circadian rhythms

As mentioned above, we intend to investigate the effect of pattern formation of peripheral circadian rhythms on the occurrence of cancer. For this reason, the mathematical models of rapid synchronization of the entire community of oscillators developed for SCN cells similar those in [33, 34, 35], are not very suitable for our purposes. In this paper we use two simple models of circadian oscillations. The first is a single-gene auto-repressor model with dimerization where the negative feedback loop is delayed in time suggested in [23]. The second is a two-variable model with both negative and positive feedback loops [22], which is a simplified version of time-delayed models proposed earlier [47, 48] for Neurospora. Both models allow not only a full synchronization of spatially distributed oscillators, but also demonstrate the pattern formation at a small enough coupling between oscillators.

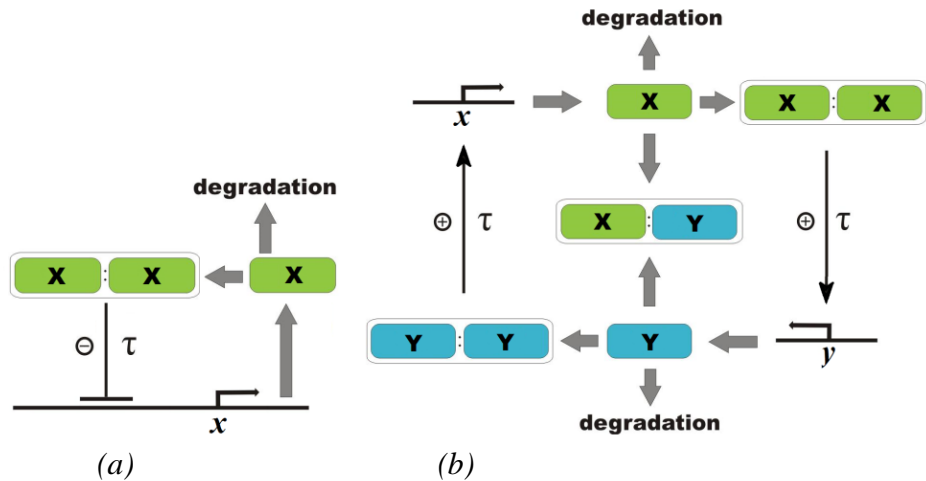

*(a)* *(b)*

Figure 3: Network architecture of the molecular components of the circadian rhythm: *(a)* - Single-gene auto-repressor model with time-delay where only $x$ gene defines the oscillatory activity. $X$ stands for the concentration of protein produced by $x$ gene; *(b)* - Two-genes model with positive and negative feedback loops where there is a pair of the principal genes, $x$ and $y$, which are responsible for maintaining the rhythm in cell. See text for details, $\tau$ symbol represents steps with a delay, $X$ and $Y$ stand for corresponding protein concentrations

### 2.3.1 Circadian model I: Single-gene auto-repressor with time-delay

The classical Goodwin model is considered to be a minimal oscillator based on a negative feedback loop [49]. It includes three-step chain of reactions for the clock gene mRNA, the clock protein and the transcriptional repressor which inhibits the production of the first component. The problem is that no limit-cycle oscillations can occur in this model until the degree of a Hill term becomes greater than 8 [50]. However, a generalization of the

Goodwin model to the case of time-delayed feedback is much more practical because the oscillatory behaviour can be obtained for the smaller value of a Hill coefficient [33]. Time-delay seems to be the most common cause of oscillations in genetic systems, since gene regulation processes are typically very slow and comprise multistage biochemical reactions engaging the sequential assembly of long molecules, and therefore are likely to generate time delays.

In fact, a three-stage scheme of the Goodwin model can be further simplified, if we take into account that the delay may also include other intermediate steps of the process. Let us consider a single-gene protein synthesis with negative auto-regulation Fig. 3a. This is a popular motif in genetic regulatory circuits, and its temporal dynamics has been analyzed within both deterministic and stochastic framework [51]. The generalized version of this system taking into account that the production of the auto-repressor protein takes a finite amount of the delay time has been studied in [23, 24].

Suppose that the protein which is responsible for the circadian rhythms can exist both in the form of isolated monomers X and dimers D. Both forms are actively interact with each other via the reactions dimerization and undimerization:

$$X + X \xrightarrow{k_{+d}} D, \qquad D \xrightarrow{k_{-d}} X + X, \qquad (6)$$

where $k_{+d}, k_{-d}$ stand for reaction rates.

We denote the unoccupied and occupied state of the promoter site of the gene as $S_0$ and $S_1$ respectively. Let us postulate that the chemical state of the operator sites $S \in \{S_0, S_1\}$ determines the production of corresponding protein at time $t + \tau$. If the operator at time $t$ is unoccupied ($S_0$) then the protein may be produced at time $t + \tau$. Otherwise, if the operator is occupied ($S_1$), the production at time $t + \tau$ is blocked. The transitions between operator states for each protein occur with rates $k_{+1}$, $k_{-1}$ when some dimer binds to the promoter or unbinds from it respectively:

$$S_0 + D \xrightarrow{k_{+1}} S_1, \qquad S_1 \xrightarrow{k_{-1}} S_0 + D. \qquad (7)$$

An important role in this model comes from the delay $\tau$ in the synthesis:

$$S_0^t \xrightarrow{A_I} S_0^{t+\tau} + X^{t+\tau}, \qquad (8)$$

where $A_I$ is the reaction rate of a time-delayed production. Finally, the system should be supplemented by the effect of protein degradation with rate $B$:

$$X \xrightarrow{B} \varnothing. \qquad (9)$$

Altogether, Eqs. (6-9) define the gene regulation at a single cell level.

The dynamics of the system (6-9) is governed by the following set of delay-differential equations (DDEs):

$$\frac{dX(t)}{dt} = -2k_{+d}X^2(t) + 2k_{-d}D(t) + A_I S_0(t - \tau) - BX(t),$$

$$\frac{dD(t)}{dt} = 2k_{+d}X^2(t) - 2k_{-d}D(t) - k_{+1}S_0(t)D(t) + k_{-1}S_1(t),$$

$$\frac{dS_0(t)}{dt} = -k_{+1}D(t)S_0(t) + k_{-1}S_1(t), \quad (10)$$

$$\frac{dS_1(t)}{dt} = k_{+1}D(t)S_0(t) - k_{-1}S_1(t),$$

where the average number of monomers and dimers at time $t$ are denoted as $X(t)$ and $D(t)$ respectively. The continuous variables $S_0(t)$ and $S_1(t)$ stand for the average number of unoccupied and occupied operator sites of clock gene at time $t$ respectively. Then we assume that

$$S_0 + S_1 = 1, \quad (11)$$

where the copy number which indicates how many copies of monomers are produced in a single act of transcription/ translation is assumed to be 1.

The main approximation we make here is an assumption that the reactions of dimerization (6) and binding/unbinding (7) are fast in comparison with production/degradation (8-9), i.e. $k_i \gg A_I, B$. Thus, we can suppose that dynamics of operator-site and dimers quickly enters into a local equilibrium, where the concentrations of reagents become

$$D = \varepsilon X^2, \quad S_1 = \varepsilon \delta S_0 X^2, \quad (12)$$

where $\varepsilon \equiv k_{+d}/k_{-d}$, $\delta \equiv k_{+1}/k_{-1}$.

We introduce the total number of the folded monomers in all forms as

$$X^T = X + 2D + 2S_1 = X + 2\varepsilon X^2 + 2\frac{\varepsilon \delta X^2}{1 + \varepsilon \delta X^2}, \quad (13)$$

where it was taken into account the Eqs. (11-12). In fact, the last term in (13) is sufficiently small compared to the other terms if $\varepsilon \ll 1$ and $\delta \ll 1$.

By taking into account (11-13), the system (10) can be reduced to the following single equation for the total number of monomers:

$$\frac{dX^T(t)}{dt} = \frac{A_I}{1 + \varepsilon \delta X^2(t - \tau)} - BX(t). \quad (14)$$

In fact, this equation is a further simplification of the Goodwin model because intermediate steps of multi-stage reaction with the formation of mRNA and transcriptional repressor have been hidden here in the delay term. A typical time series of the number of the X monomers obtained by solving Eqs. (14) for a single cell is shown in Fig. 4a. For the parameter values given in the figure caption the period of oscillations is about 27 h. This is about three times greater than the time delay $\tau$.

In order to describe the effect of intercellular signalling, we suggest for simplicity that the monomer can penetrate to other cell via the mechanism of the diffusive transport. Thus, we can generalize Eq. (14) for the case of a multi-cellular system with the circadian signalling in the following way:

$$\frac{dX_i(t)}{dt} = \frac{1}{1 + 4\varepsilon X_i(t)}\left(\frac{A_I}{1 + \varepsilon \delta X_i^2(t - \tau)} - BX_i(t)\right) +$$
$$+ \alpha \sum_{j \in adj(i)} P_{ij}(X_j(t) - X_i(t)), \quad (15)$$

where the subscript "$i$" refers to the $i$-cell. The prefactor has appeared in Eq. (15) due to formula (13) where the number of dimers binding to the promoter has been neglected. The last term in Eq. (15) describes the transport of the signalling species between neighbouring cells $i$ and $j$ ($adj(i)$ stands for "adjacent to $i$-cell"), with the flux being proportional to the boundary length $P_{ij}$ and tuned by the coefficient $\alpha$. This implies that the transport rate is limited by the transfer through cell membranes. After a cell divides, the daughter cells inherit the phase of the circadian rhythm of the parent cell. We assume for the simplicity that the same X protein transports the circadian signal outside the cell.

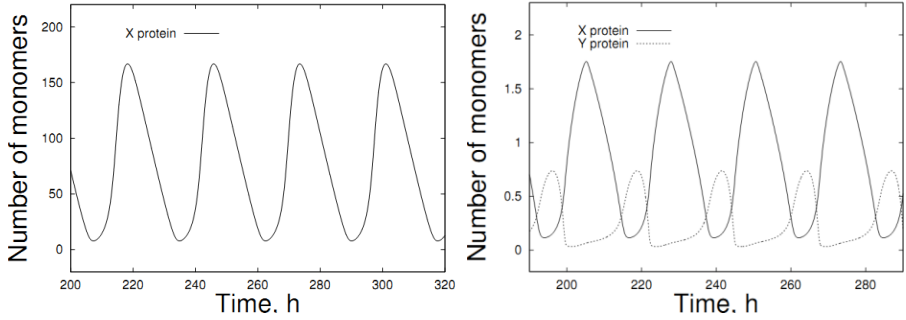

(a) (b)

Figure 4: The typical time series obtained during a single-cell simulation: (a) - Dynamics of the X protein within the Circadian Model I defined by Eq. (14) with parameters $\tau = 8$ h, $A_I = 5000$ nM h$^{-1}$, $B = 5$ h$^{-1}$, $\varepsilon = 0.1$ nM$^{-1}$, $\delta = 0.2$ nM$^{-1}$; (b) - Dynamics of the X (solid line) and Y (dashed line) proteins within the Circadian Model II defined by Eqs. (26,27) with parameters $\tau = 6$ h, $A_X = 8$ nM h$^{-1}$, $A_Y = 4$ nM h$^{-1}$, $B_X = 0.3$ h$^{-1}$, $B_Y = 0.4$ h$^{-1}$, $\varepsilon_i = 5$ nM$^{-1}$, $\delta_i = 5$ nM$^{-1}$

At a single-cell level without the intercellular signalling ($\alpha = 0$) the equation (14) has only a positive steady-state solution:

$$X^* = -\frac{B}{2A_I \varepsilon \delta} + \sqrt{\frac{B^2}{4A_I^2 \varepsilon^2 \delta^2} + \frac{1}{\varepsilon \delta}}. \quad (16)$$

By linearizing Eq. (14) near the fixed point (16), and looking for a solution of the form $X(t) : e^{\lambda t}$, where $\lambda = \chi + i\omega$, we obtain

$$\chi = \frac{1}{\tau} Re(W(-2\tau A_I \varepsilon \delta x^* e^{\tau B})) - B, \quad (17)$$

$$\omega = \frac{1}{\tau} Im(W(-2\tau A_I \varepsilon \delta x^* e^{\tau B})), \quad (18)$$

where $W(z)$ stands for the Lambert function defined as $W(z)e^{W(z)} = z$. The condition for the vanishing of the function (17) gives the analytical expression for the Hopf bifurcation curve, and the frequency of neutral oscillations is defined by (18). The big advantage of the models with delay is the weak dependence of the oscillation period on the reaction rates of genetic reactions. This is especially important when describing the circadian rhythms, as it significantly increases the robustness of the model. The period of oscillations in the models with a strong delay heavily depends on the delay time. Mathematically, it can be explained by the asymptotic behaviour of the Lambert function in (18).

### 2.3.2 Circadian model II: Two-genes oscillator with positive and negative feedback loops

Figure 3b shows a graphical scheme of the general network architecture. The genes are labeled as $x$ and $y$, and their proteins, as $X$ and $Y$, respectively. They constitute a system having both activating and repressing elements [22]. One can see from Fig. 3b that the protein $Y$ dimerises and then works as a positive regulator of X by initiating its transcription. The $X$ operates symmetrically with respect to the $Y$. During the circadian cycle the proteins are degraded by turns which enables the cycle to restart again and again. The scheme also takes into account that the $X$ can be also removed by associating with $Y$ to form a heterodimeric X/Y complex. The synthesis of both $X$ and $Y$ occurs with a delay of several hours (Fig. 3b).

The molecular mechanism described above can be presented as a set of biochemical reactions:

$$X + X \xrightarrow{k^X_{+d}} D_X, \qquad Y + Y \xrightarrow{k^Y_{+d}} D_Y, \tag{19}$$

$$D_X \xrightarrow{k^X_{-d}} X + X, \qquad D_Y \xrightarrow{k^Y_{-d}} Y + Y, \tag{20}$$

$$S_0^X + D_Y \xrightarrow{k^X_{+1}} S_1^X, \qquad S_0^Y + D_X \xrightarrow{k^Y_{+1}} S_1^Y, \tag{21}$$

$$S_1^X \xrightarrow{k^X_{-1}} S_0^X + D_Y, \qquad S_1^Y \xrightarrow{k^Y_{-1}} S_0^Y + D_X, \tag{22}$$

$$S_1^X \xrightarrow{A_X} S_1^X + X^{t+\tau}, \qquad S_1^Y \xrightarrow{A_Y} S_1^Y + Y^{t+\tau}, \tag{23}$$

$$X \xrightarrow{B_X} \varnothing, \qquad Y \xrightarrow{B_Y} \varnothing, \tag{24}$$

$$X + Y \xrightarrow{k} \varnothing. \tag{25}$$

Here, the reactions of dimerisation and dedimerisation are given by Eqs. (19) and (20); the dynamics of operator-sites during binding/unbinding of dimers are described by Eqs. (21) and (22); the time-delayed synthesis of proteins is given by Eq. (23); the linear and non-linear degradation are described by Eqs. (24) and (25), respectively. Assuming that reactions of dimerisation and binding/unbinding are fast compared to the processes of production and degradation of proteins, so that they quickly reach a local equilibrium, we obtain two delay differential equations (in fact, the procedure of derivation is similar to those for the Circadian Model I):

$$\frac{dX^T(t)}{dt} = A_X - \frac{A_X}{1 + \varepsilon_Y \delta_X Y^2(t-\tau)} - B_X X(t) - kX(t)Y(t), \tag{26}$$

$$\frac{dY^T(t)}{dt} = A_Y - \frac{A_Y}{1 + \varepsilon_X \delta_Y X^2(t-\tau)} - B_Y Y(t) - kX(t)Y(t), \tag{27}$$

where $\varepsilon_X \equiv k_{+d}^X/k_{-d}^X$, $\delta_X \equiv k_{+1}^X/k_{-1}^X$, $\varepsilon_Y \equiv k_{+d}^Y/k_{-d}^Y$, $\delta_Y \equiv k_{+1}^Y/k_{-1}^Y$ are equilibrium constants of the fast stages. The average number of the X and Y monomers at time $t$ are denoted as $X(t)$ and $Y(t)$ respectively. The $X^T$ and $Y^T$ are the total number of the folded X and Y monomers in all forms defined in the same way as it was done in (13). For simplicity, the delay time $\tau$ is assumed to be same for both genes. In fact, the delay in the model (26,27) includes all possible multi-stage events between the moment of the transcription start and the moment when the protein becomes able for regulation. For example, the delay in Eq. (26) includes Y complexation and nuclear entry of the complex, X translation and multiple post-transcriptional phosphorylations. The delay in Eq. (27) is the time taken for X protein to activate $y$-gene and the production of Y. It includes X nuclear entry, Y translation and possible multiple phosphorylation of Y. Thus, the phosphorylation reactions of X and X/Y complexes are not taken into consideration explicitly and are lumped in the delays because the number of X phosphorylations are not definitely known, but only appears to be high. This is the difference with 23-variable model suggested in [47]. The same way was chosen in the paper [48] (see the discussion in reference).

A typical time series of the total number of the X and Y monomers obtained by solving Eqs. (26,27) for a single cell is shown in Fig. 4b. One can see that the X protein dominates in the "day time", reaching its maximum at "noon" when the concentration of the Y protein is close to zero. During the "night", the situation is opposite: $X \to 0$, and Y is maximal.

As before, we can generalize Eqs. (26,27) for the multi-cellular case with the circadian signalling between cells in the following way:

$$dX_i(t)dt = \frac{1}{1+4\varepsilon_X X_i(t)}\left(\frac{A_X \varepsilon_Y \delta_X Y_i^2(t-\tau)}{1+\varepsilon_Y \delta_X Y_i^2(t-\tau)} - B_X X_i(t) - kX_i(t)Y_i(t)\right) + \quad (28)$$
$$+ \alpha \sum_{j \in adj(i)} P_{ij}(X_j(t) - X_i(t)),$$

$$dY_i(t)dt = \frac{1}{1+4\varepsilon_Y Y_i(t)}\left(\frac{A_Y \varepsilon_X \delta_Y X_i^2(t-\tau)}{1+\varepsilon_X \delta_Y X_i^2(t-\tau)} - B_Y Y_i(t) - kX_i(t)Y_i(t)\right), \quad (29)$$

where the subscript "$i$" refers to the $i$-cell and $X_i$ and $X_i$ are defined as specific quantities (per square unit) and are set pointwise in each cell. We assume that the X protein transports the circadian signal outside the cell, while its partner Y is confined within the cell volume. Similar dynamic equations were studied earlier in the framework of a continuous spatially-extended reaction-diffusion model [22].

In fact, the intracellular circadian clock can be generally affected by a cell size change. The robustness of the oscillations in this case was studied recently in [52] within the stochastic description. Since we work here in terms of the concentration, rather than the number of molecules as it is commonly used in the stochastic models, this possible dependence is not taken into account.

## 2.4 Submodel of transformation of cells

As mentioned above, the main circadian genes appear to greatly influence the tumourigenesis. As is known, the rhythmicity demonstrated in the expression of clock-controlled genes regulates various functions of cells, such as their division and proliferation (see, for example, theoretical papers [53, 54] and experimental observations [55]). Desynchronisation of this rhythmicity can be involved in some pathologies, including the development of tumours. There is increasing evidence linking malfunction of the bioclockwork and pathogenesis of cancer [18, 19, 20, 27, 28, 56, 57]. In the review [28], it was given the large number of examples of the connection between circadian genes, circadian periodicity, aging-related phenotypes, and cancer. For instance, it was shown experimentally that mice homozygous for disruption of circadian clock genes that result in loss of rhythmicity have many phenotypic anomalies. Among these phenotypes are increases in basal cancer occurrence or in the occurrence of cancer following genotoxic stress or in genetically cancer-prone models [56]. In another review [57], it was discussed the emerging role of circadian genes in hematological and hormone-related malignances and even was declared that "manipulating circadian biology is a possible way to fight cancer".

But what is the exact mechanism linking the circadian rhythmicity and the occurrence of cancer? Some details of this connection began to appear in the literature. For example, it was shown recently that transcriptional regulator at the core of the circadian clock mechanism hPer2 directly affects hp53 levels which plays a key role as human tumour suppressor [42]. The important result obtained experimentally was reported in [58]. The authors have checked whether the circadian rhythm was connected to the cell-cycle oscillation in immortalized $rat-1$ fibroblasts by observing cell-cycle gene promoter-driven luciferase activity. It was revealed that there was no direct phase relationship between the circadian and cell rhythms. These data imply that the circadian system does not govern the cell-mitosis rhythm in $rat-1$ fibroblasts. These experimental results rather differ from numerous studies in which assume a priori that cell mitosis is regulated by circadian system in mammalian tissues in vivo. Thus, it was suggested that there is no direct coupling between the circadian rhythm and cell cycle but the timing of cell mitosis is synchronized with the rhythmic host environment [58]. Another study implying that the coupling among cell population can plays the role in preventing the carcinoma is [43] where authors theoretically show that the coupling between neighboring cells significantly improves the sustainability of hp53 pulses. Since the exact mechanism linking circadian oscillations with cancer is still unknown, we propose a simple phenomenological model. Based on the papers cited above, the main idea of the alteration mechanism is a synchronisation failure of a local oscillation phase in the common field of spatially synchronised circadian rhythms in the epithelial tissue.

Preliminary numerical simulations of the spatially extended problems within the circadian model I (15) and model II (28-29) have shown that when the number of cells is large enough, a complete synchronisation, *i.e.* the total alignment of the oscillation phases in all cells, cannot be achieved when the coupling is small (see the numerical examples in the next Section). Instead, the cells are organised in collective spatio-temporal patterns including clusters of cells oscillating with almost the same phase. The thin layer of cells between clusters exhibit oscillations with intermediate phase. Thus, we introduce a new variable which characterises the local dephasing defined as the phase difference of the circadian rhythm of an $i$ th cell and the average phase of the adjacent cells:

$$\Phi_i(t) = <|\phi_i(t) - \phi_k(t)|>_{k \in adj(i)}, \tag{30}$$

The oscillation phase $\phi$ can be defined in different ways but we choose a physical rather than formal definition and characterise it by the concentration difference

$$Model \quad I: \quad \phi_i^I(t) = \frac{X_i(t)}{\max(X_i)}, \tag{31}$$

$$Model \quad II: \quad \phi_i^{II}(t) = \frac{X_i(t) - Y_i(t)}{\max(X_i, Y_i)}, \tag{32}$$

where both expressions are normalized by its maximal value. If the $i$-th cell is in a fully synchronised field, the value of $\Phi_i$, obviously, will be zero. On the other hand, the dephasing (30) increases near the boundary of the clusters. The maximum value of the dephasing would occur in case of an isolated cell within a phase cluster. According to our hypothesis, this cell is most at risk to become a cancer cell. Apparently, in a dynamically changing pattern where cluster boundaries are varying in time, the cell transformation should occur much less frequently.

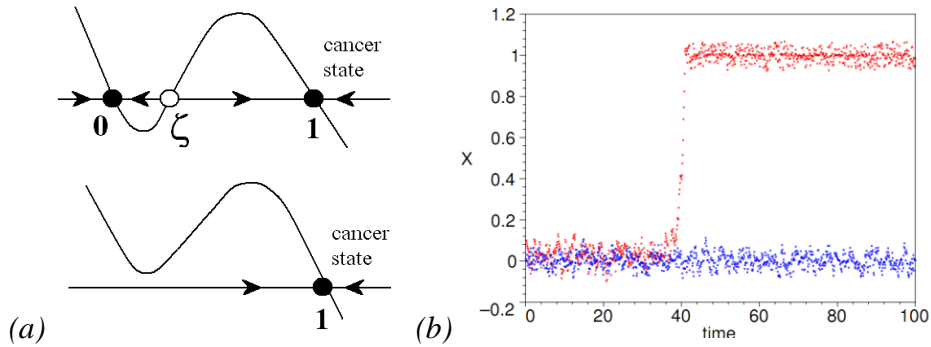

Figure 5: *(a)* The phase portrait of Eq. (33) without noise (top) and with a non-zero stochastic input (bottom); *(b)* A time series for a normal cell (blue) and a cell transformating into the cancerous state (red).

At this stage, we need to determine more accurately what we mean by cancer in our work. Generally speaking, the metamorphosis of healthy cell into cancer looks like a chain reaction caused by initial errors, which result in more severe errors, each increasingly enabling the cell to avoid the controls that limit normal tissue growth. This implies that the cell undergoes several transformations in sequence, until it finally becomes a full-fledged cancer cell, part of the invasive tumour. Since in this paper we focus on the dynamics of a large number of cells entering into the chemo-mechanical interaction, we restrict ourselves to the simultaneous transformation of a healthy cell into a cancer cell, which allows the use of simple model of bistability.

Bistability is known to be important phenomenon for understanding main events of cellular functioning, such as decision-making processes during cell cycle, differentiation, apoptosis and so on. This concept is also extensively used to describe the loss of cellular homeostasis associated with early development of cancer [59]. More importantly, the behaviour of bistable systems exhibit hysteresis. It means it can generate different output values for the same input value depending on its state, a key property for switch-like control of cellular processes. This is what we need in our case.

Under these assumptions, the equation for the state of $i$ th cell can be written as

$$\frac{dZ_i}{dt} = -\lambda Z_i(1-Z_i)(\zeta - Z_i) + N\Phi_i \xi_i(t), \tag{33}$$

where $Z$ is the state function, $\lambda$ is the damping parameter, $\xi(t)$ is an uncorrelated zero mean noise input in the range $\xi \in [-1;1]$, $N$ is the amplitude of the noise, assumed to be multiplicative. Without noise, Eq. (33) has two stable stationary solutions: $Z = 0$, standing for the normal state of the cell, and $Z = 1$ corresponding to the cancer state (indicated in Fig. 5a by black circles). These solutions are separated by an unstable steady state which separates their domains of attraction (open circle in Fig. 5a).

The parameters of the model (see Table 1) are calibrated in such a way that transition would occur under the influence of noise, but only at a high value of the dephasing (30), and the reverse transition of already transformed cell is impossible (Fig. 5b).

Table 1: List of principal parameters. Spatial dimensions and force are measured in terms of arbitrary units L and F respectively.

| | |
|---|---|
| $\mu = 1.0$ FL$^{-1}$ | $K = 1.0$ LF$^{-1}$h$^{-1}$ |
| $\eta = 1.0$ FL$^{-3}$ | $F_0 = 0.02$ F |
| $p_0 = 2 \times 10^{-4}$ | $l_0 = 0.15$ L |
| $q = 1.4$ | $\lambda = 10$ h$^{-1}$ |
| $\alpha = 0.1$ L$^{-1}$h$^{-1}$ | $\zeta = 0.15$ |
| $\tau = 6.0$ h | $A_0 = 3\sqrt{3}/2$ L$^2$ |
| $\varepsilon_X = 5.0$ nM$^{-1}$ | $\varepsilon_Y = 5.0$ nM$^{-1}$ |
| $\delta_X = 5.0$ nM$^{-1}$ | $\delta_Y = 5.0$ nM$^{-1}$ |
| $k = 30$ nM$^{-1}$h$^{-1}$ | $B_X = 0.3$ h$^{-1}$ |
| $A_X = 8.0$ nM h$^{-1}$ | $B_Y = 0.4$ h$^{-1}$ |
| $A_Y = 4.0$ nM h$^{-1}$ | $N = 2.5$ h$^{-1}$ |

# 3 Numerical results

## 3.1 Numerical methods

The initial configuration of the system is a regular hexagonal lattice comprising 1560 healthy cells and no cancer cells. The shape and location of each cell is defined by its nodes. The tissue as a whole has the form of a stripe with two free borders with periodic boundary conditions applied there. The typical values of the parameters for normal cells are given in Table 1. Most of the parameter values of mechanical submodel are taken from our previous work [21], where the values were calibrated so as to give the realistic results. Parameter values for circadian rhythms submodel have been taken from [47, 48], plus we add some our own, e.g. for dimerization and diffusion.

The set of differential equations describing the dynamics of mechanical subsystem (1-5), the circadian rhythms and signalling ((15) or (28-29) depending on the circadian model) and the state function dynamics (30-33) has been solved using the explicit Euler method, whose stability was warranted by a sufficiently small time step $\Delta t = 0.005$. The time step for the calculation of the molecular processes in cells was synchronised with the step of calculating the mechanical movement of the tissue cells. In order to avoid storing a large amount of data for long time delay intervals, we have used a novel numerical simulation algorithm [22] based on adaptive storing in the computer memory of some selected nodal data only, and a subsequent interpolation to calculate the intermediate values. This adaptive numerical scheme for data storing can speed up the calculation up to 8 times without visible deviations from benchmark simulation.

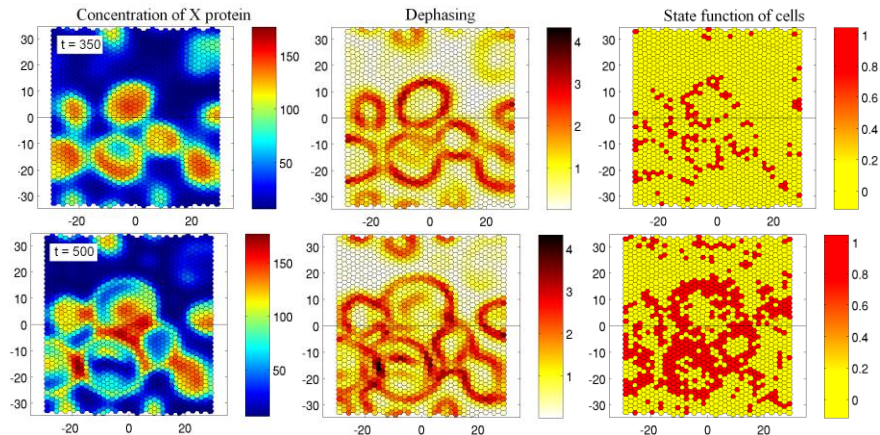

Figure 6: Evolution of the X protein concentration (first column), the dephasing $\Phi$ (second column) and the state function $Z$ (third column) in the epithelium composed by 1560 cells calculated for the Circadian model I with transformation parameters $\lambda = 10$ h$^{-1}$, $\zeta = 0.15$, $N = 0.1$ h$^{-1}$. The evolution of the system starts from random phase distribution. The upper and lower rows of frames correspond to $t = 350$ h and 500 h respectively.

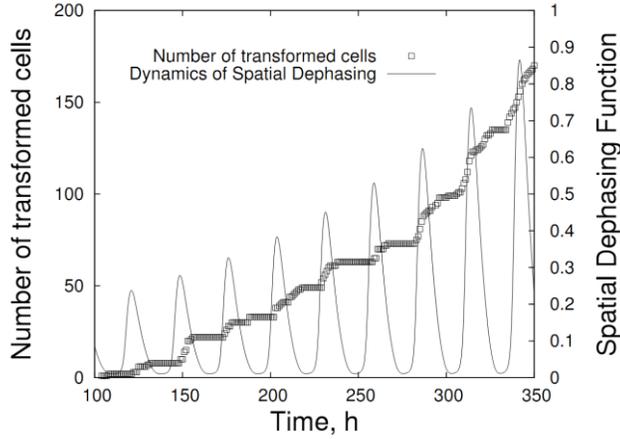

Figure 7: The number of transformed cells (left axis) and spatial dephasing function $R(t)$ (right axis) as a function of time calculated for the Circadian model I with transformation parameters $\lambda = 10$ h$^{-1}$, $\zeta = 0.15$, $N = 0.1$ h$^{-1}$

## 3.2 Synchronisation of circadian rhythms

### 3.2.1 Circadian model I

We take as the initial condition a random phase distribution. In the process of evolution, the rhythms in cells try to synchronise through a weak nonlinear interaction given by the last term in Eqs. (15), generating the various spatio-temporal patterns. From the perspective of biology, an arbitrary distribution of phases in cells taken as the initial conditions looks artificial, since synchronisation of rhythms happens at the stage of embryonic development. Nevertheless, the numerical study with random initial conditions allows us to better understand the self-organisation properties of the system and to evaluate the mechanisms of pattern formation.

The typical values of the parameters governing the circadian rhythms in each cell within the Model I are as follows: $A_I = 5000$ nM h$^{-1}$, $B = 5$ h$^{-1}$, $\varepsilon = 0.1$ nM$^{-1}$, $\delta = 0.2$ nM$^{-1}$, $\tau = 8$ h (see Fig. 4a). Fig. 6 presents the results of numerical simulation of the temporal evolution of the system (15). The patterns of the concentration $x$, the dephasing $\Phi$ and the state function $Z$ are shown for two consecutive moments of time. We have found that the spatial dynamics includes the slow development of quasi-standing waves which arise against the mean field oscillating with the basic period 27 h (Fig. 6, left column). The dephasing $\Phi$ calculated according to (30,31) reaches a maximum value on the boundary of the standing waves domain (Fig. 6, central column). The cells that fall into this region, are at high risk to transform into cancerous state. The right column in Fig. 6 confirms this conclusion: most of the transformed cells are grouped in a zone of uncertain oscillation phase characterized by the maximal dephasing. The important point is the small mobility of the pattern shown in the figure. Since the boundary of the standing waves change slowly, the same cells are always at risk of the transformation. Note that the parameter values for the transformation submodel (33) have been calibrated so that the number of transformed cells was sufficient to demonstrate the effect ($\lambda = 10$ h$^{-1}$, $\zeta = 0.15$, $N = 0.1$ h$^{-1}$).

In order to characterise how the dephasing jump can speed up the cell transformation we introduce a measure given by the the dephasing $\Phi$ computed in terms of the area of the dissynchronization zone (i.e. the number of cells where $\Phi(t)$ is larger than an arbitrary threshold $\Phi_0$) normalised by the total number of cells in the system versus time. In other words, we compute as a function of time the quantity:

$$R(t) = \frac{1}{M} \sum_{i=1}^{M} \psi_i, \qquad (34)$$

where $\psi = 1$ if $\Phi(t) > \Phi_0$ and $\psi = 0$ otherwise, $M$ is the total number of cells. In fact, the spatial dephasing function $R(t)$ is the analogue of the spatial reaction rate calculated commonly in pattern formation studies in the nonlinear chemistry. The typical value we have used for the threshold $\Phi_0$ is $0.2\Phi_{max}$. The Figure 7 present the number of transformed cells (left axis) and spatial dephasing function $R(t)$ (right axis) as a function of time. It can be seen that the spatial dephasing function oscillates with the period of the basic circadian rhythm reaching a maximum value once per day. The absolute majority of irreversible cell transformations occur just at this period of time.

### 3.2.2 Circadian model II

We found that the spatio-temporal dynamics observed within the circadian model II is more varied compared with the model I. The principal dynamical mode of the system here is a travelling spiral wave pattern, similar to a related continuous system [22]. The waves arise from initial disturbances, and travel outward from the source, until this pattern spreads to the entire area. The wave front is originally well ordered, but secondary instabilities appear due to numerous secondary excitation nuclei. Nonlinear interplay between wave structures results in the development of a chaotic pattern via the mechanism of core breakup. Spiral wave patterns are most common in two-dimensional systems far from equilibrium [60, 61]. Spiral waves are observed in a wide range of biological and chemical systems, observed in liquid phase reaction-diffusion systems [62, 63], on catalytic surfaces [64], and in subcellular processes [65].

Fig. 8 presents the results of numerical simulation of the temporal evolution of the system (28,29) with the effect of cell transformation governed by Eqs. (30-33). The patterns of the concentration $x$, the dephasing $\Phi$ and the state function $Z$ are shown at the same time $t = 500$ h.

If the degradation rate is low (the upper row in Fig. 8) the scenario is as follows. After the evolution of the system starts from random phase distribution, the nonlinear dynamics demonstrates two different spatio-temporal behaviour: the mean field oscillations with the period about 22.6 h and a spiral travelling wave pattern (Fig. 8, first frame in the upper row).

Since the spiral wave moves rapidly in space, the number of cell transformation occurring due to dephasing is small (Fig. 8, second and third frames in the upper row).

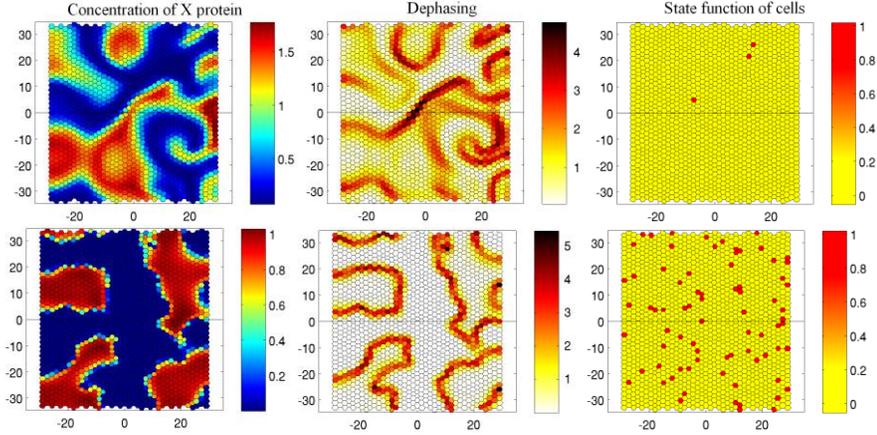

Figure 8: Concentration of the X protein (first column), the dephasing $\Phi$ (second column) and the state function $Z$ (third column) in the epithelium composed by 1560 cells calculated for the Circadian model II shown at $t = 500$ h. The evolution of the system started from random phase distribution. The frames correspond to the case of small degradation with $B_X = 0.3$ h$^{-1}$, $B_Y = 0.4$ h$^{-1}$ (the upper row) and large degradation $B_X = 6$ h$^{-1}$, $B_Y = 8$ h$^{-1}$ (the lower row). Please check the Supplementary multimedia files to see the animation.

When the rate of degradation of proteins increases sharply, we observe, instead of the travelling waves, quasi-steady clustering. The cells are self-organised in two large groups with a particular phase of oscillations separated by cells fluctuating with intermediate phases (Fig. 8, the lower row). Cluster formation in the systems with a large number of discrete elements which exchange signals has been also recently observed in interacting synthetic genetic oscillators [66]. This kind of clustering seems to be a significant feature of diverse communities. Therefore it may strongly affect differentiation of cells in the various organs. The effect of the transition to a quasi-steady clustering pattern with a sharp increase of the linear protein degradation in cells may be also responsible for ageing of tissues.

Our numerical experiments show that the clastering regime, like in the case of the circadian model I, alternation of healthy cells (Fig. 8, second and third frames in the lower row) is promoted in the stagnant zone between the clusters. Thus, a sharp rise of protein degradation and change of the spatial dynamics can be seen as the key event in the transformation process.

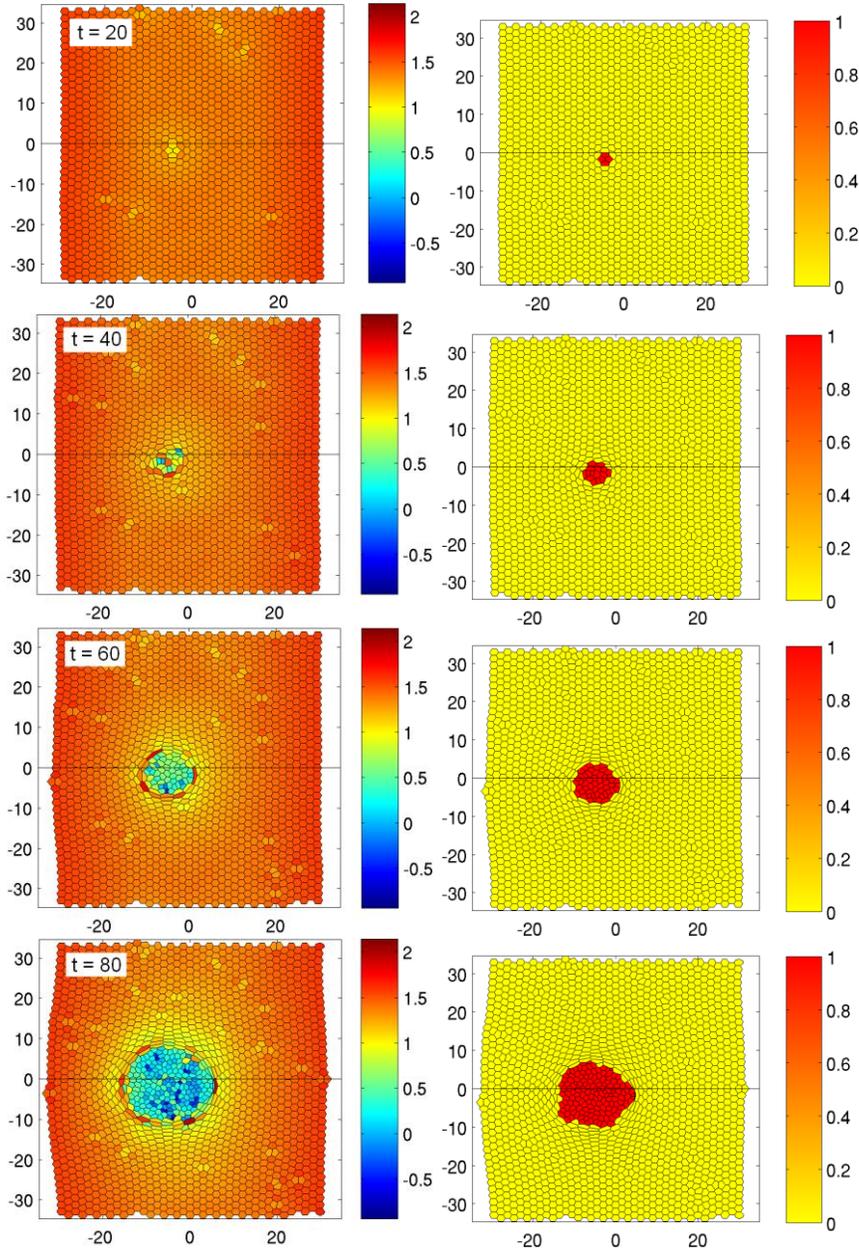

Figure 9: Evolution of a non-invasive tumour in epithelial tissue in time: the colour code shows the relative cell size with respect to the average size of a normal cell (*left*) and the state function $Z$ where $Z = 1$ and $Z = 0$ stand for cancer and normal cells, respectively (*right*). The frames arranged from top to bottom correspond to times $t = 20, 40, 60, 80$ h respectively. The parameters defining the properties of a cancer cell are $A_0 = 3\sqrt{3}/2$ L$^2$, $\mu = 1.2$ FL$^{-1}$, $\eta = 1.0$ FL$^{-3}$, $l_0 = 0.15$ L, $T = 12 \pm 2$ h. Please check the Supplementary multimedia files to see the tumour growth in dynamics.

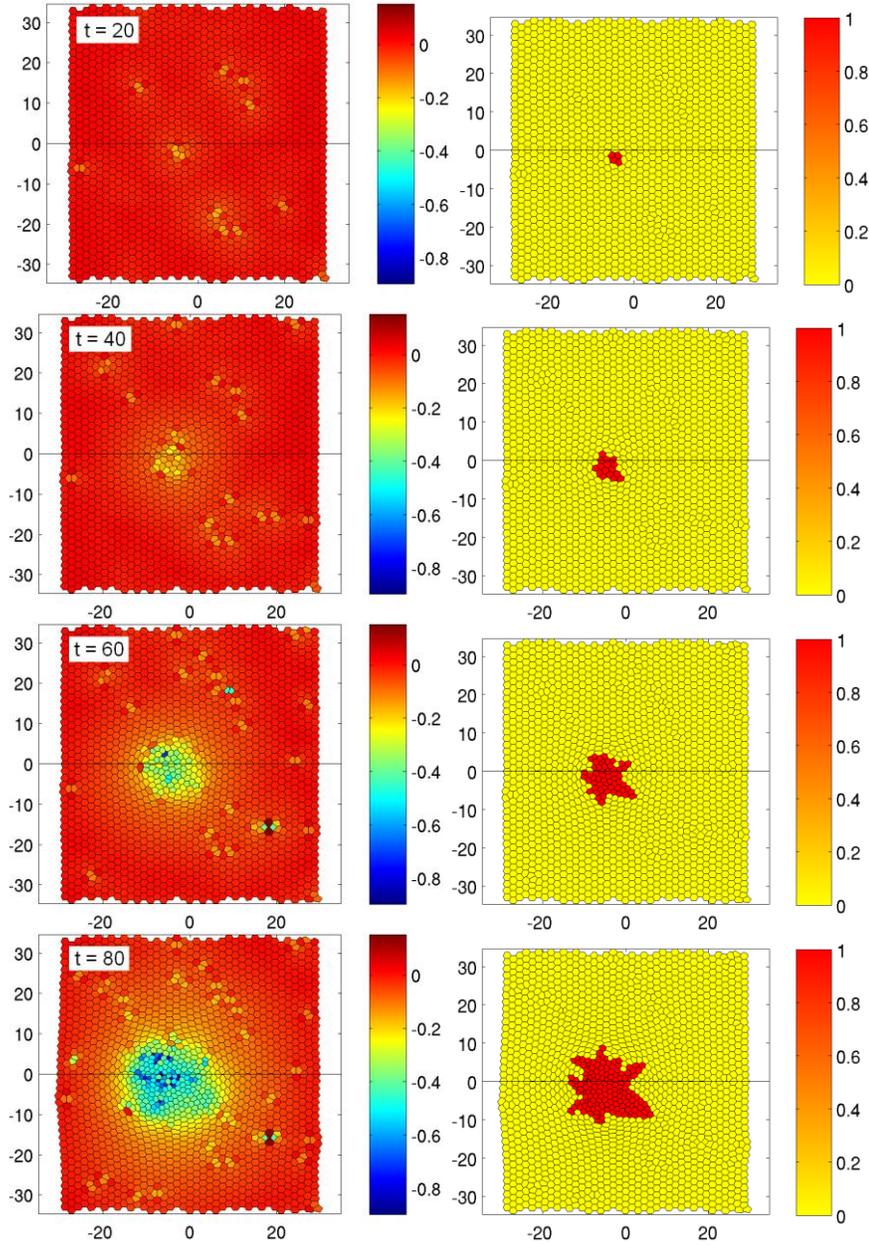

Figure 10: Evolution of an invasive cancer tumour in epithelial tissue in time: he colour code shows the relative cell size with respect to the average size of a normal cell ( *left*) and the state function $Z$ where $Z = 1$ and $Z = 0$ stand for cancer and normal cells, respectively ( *right*). The frames arranged from top to bottom correspond to times $t = 20, 40, 60, 80$ h respectively. The parameters defining the properties of a cancer cell are $A_0 = 3\sqrt{3}/2$ L$^2$, $\mu = 1.2$ FL$^{-1}$, $\eta = 1.0$ FL$^{-3}$, $l_0 = 0.4$ L, $T = 12 \pm 2$ h. Please check the Supplementary multimedia files to see the tumour growth in dynamics.

## 3.3 Tumour development

The entire set of information available to date about the cancer cells suggests that their chemo-mechanical and biological properties are substantially modified compared to normal cells. The most important differences have been underlined in the Introduction. While a regular cell stops dividing if some genetic alternation occurs, cancer cell continues to divide. As a result of continued cell division, many daughter cells with abnormal DNA emerge. They constitute a separate species of cells that form their own medium – tumour. To model this transition, we have introduced two distinct set of parameters characterising the physical and chemo-mechanical properties of the two type of cells, normal and cancerous.

The proposed model includes the following special properties of cancer cells:

- the averaged size of cancer cells is governed by the same reference cell area $A_0$ (cancer cells can be either larger or smaller normal tissue cells);

- changes of the perimeter and the area of cancer cells are governed by different values of elasticities $\mu$ and $\eta$;

- the mode of a cancer cell division does not follow the probability relation (4) depending on the shape of the cell, but is forced instead to a specific period of division $T$ with a weak scattering;

- the mechanism of cell transformation prevents the reverse transition of cancer cells into normal state;

- the high mobility (invasivity) of cancer cells is set by the parameter $l_0$, the higher values of which initiate the process of intensive intercalation in the ensemble of cancer cells;

- since the circadian genes in cancer cells mutate, and the circadian rhythm does not work properly, the circadian dynamics defining the rhythm is suspended in cancer cells;

- the cancer cells do not exchange signals with normal cells and do not participate in the synchronisation of the circadian rhythm throughout tissue.

At the same time, the cancer cells continue to interact mechanically with tissue, and, in turn, put pressure on the healthy cells. It is natural that each descendant of cancer cell inherits all properties of its parent.

Figures 9 and 10 present numerical simulations of two different types of cancer. Fig. 9 shows the evolution of a non-invasive tumour consisting of fairly large transformed cells. The parameters defining the properties of cancer cells are $A_0 = 3\sqrt{3}/2$ $L^2$, $\mu = 1.2$ $FL^{-1}$, $\eta = 1.0$ $FL^{-3}$, $l_0 = 0.15$ L, $T = 12 \pm 2$ L. In order to focus upon the development of the tumour, the alteration process has been stopped during the simulation run just after the first cell turned into a cancer cell. This moment of time was fixed to $t = 0$ h.

One can see that the size and shape of the cells becoming cancerous differ from those of the normal members of the community. Cancer cells are approximately twice as large and irregular in shape. The healthy cells that border on cancer cells experience a significant stress: they are squeezed and shrunk under the onslaught of cancer cells. Since the period of the division of cancer cells is shorter, the tumour evolves rapidly, increasing the occupied area. The tumour remains in this case roughly round with the front between healthy and cancerous cells being relatively stable (Fig. 9). It is interesting to note the internal structure of the tumour itself – it looks like a cyst with walls lined with relatively large cells, while the internal space consists of smaller cells. The forced proliferation of cancer cells is the reason why in the bulk of tumour the cells are also strongly squeezed and may become irregularly shaped.

Another type of tumour development is shown in Fig. 10. Here one can see the evolution of a vigorously invasive carcinoma, consisting of relatively small cells. The parameters defining the properties of cancer cells are $A_0 = 3\sqrt{3}/2$ $L^2$, $\mu = 1.2$ $FL^{-1}$, $\eta = 1.0$ $FL^{-3}$, $l_0 = 0.4$ L, $T = 12 \pm 2$ h. The main difference with the previous case is that the threshold of intercalation $l_0$ for cancer cells has been sharply increased in comparison with the normal cells. This condition allows to reduce a high level of potential energy in cells by restructuring their form due to the intercalation process. The characteristic rate of cell intercalation exceeds here the division frequency, and the front between the tumour and the healthy tissue becomes unstable. Because of the easy intercalation, active proliferation and relatively small size, the cancer cells actively change their location, move apart and migrate through the healthy tissue. Thus, this simulation models a key property of the invasive type of cancer being capable to penetrate into various tissues and organs.

The time evolution of the radial positions of all tumour cells is presented in Fig. 11. It is clearly seen that the pathways of cancer cells often overlap, reflecting the active intercalation process. One can also notice that the "center of gravity" of the tumour, generally, also moves, slowly departing from the location of the starting alteration event. The instability of the front between normal and cancer cells, clearly visible in Fig. 10, looks similar to a fingering instability at the interface of two immiscible fluids [67] when a fluid with a lower viscosity is pushed into a fluid of higher viscosity. Although tissue is neither a liquid nor a granular medium, and cannot be characterised as either miscible or immiscible, the analogy is natural, as the effective viscosity of the tumour is lowered by easy intercalation, while the driving pressure is generated by division of cancerous cells.

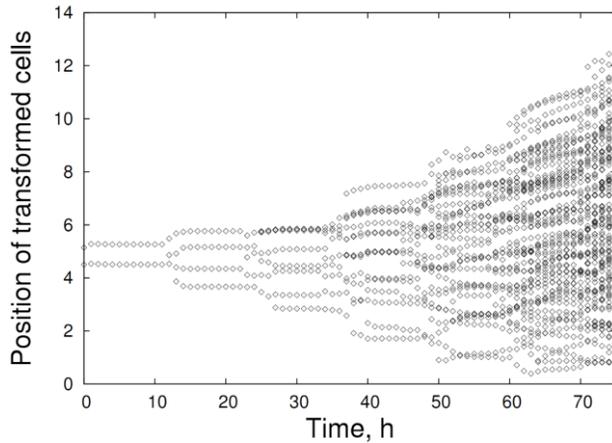

Figure 11: Time series of radius positions of cancer cells for the invasive carcinoma shown in Fig. 10, demonstrating the tumour proliferation and the intercalation of cancer cells

## 4 Conclusions and outlook

Cancer formation is certainly a complex biophysical process, and therefore the cancer modelling necessitates a miltiscale mathematical approach. We propose a minimal multiscale chemo-mechanical model which includes three natural scales of the tumour formation. Processes on the mesoscopic (cellular) scale are activated by circadian rhythm signals generated on the misroscopic (subcellular) scale, and exert an influence over cooperative phenomena occurring at the macroscopic level characterised by a long-range coordination of cells. The medium where the tumour grows and spreads is modelled here taking into account both transport of signalling species and mechanical interactions between the cells. Many details of this model still need experimental verification.

This paper contains several new results. Two simple schemes have been suggested to describe the circadian rhythms: one is a single-gene auto-repressor model and the other is a two-variable model with dimerization. The main difference between these models is that the first model includes only a negative feedback loop, while the second one in addition has a strong positive feedback loop. Positive feedback is a process which makes the input larger and therefore it tends to lead the system to instability. In contrast, a system with negative feedback tends to reduce the perturbation. Thus, the concept of positive feedback is closely related to the concept of active continuous medium in which energy is pumped from the medium into the propagating wave. Since the elements of positive feedback have been reliably found in the molecular network architecture of the circadian rhythms for some simple organisms, the Circadian Model II seems to be more realistic. Although the Circadian Model I gives an example of the simplest oscillating system. As a consequence, if the Model I describes the spatial dynamics in the form of quasi-standing waves, the Model II can demonstrate the development of vigorous travelling waves (in the case when the positive feedback is stronger than negative one). We have shown within our transformation submodel that the alternation process of healthy cell is promoted when the negative feedback dominates because the pattern formation here becomes more steady and stagnant. Thus, the occurrence of cancer cell may be associated with a local failure in the circadian rhythms caused by the features of a synchronized protein pattern in the tissue. Then we have demonstrated numerically that the invasive form of cancer may occur due to the physical properties of cancer cells, which have a higher mobility in tissue: the interface between the tumour and healthy tissue is experiencing a wave instability due to collective chemo-mechanical interaction of cells.

We believe that biologists turn to spatially distributed experimental data would gradually open up opportunities for modelling spatially distributed biochemical processes. Such spatio-temporal experimental data on the circadian rhythms obtained using fluorescent reporters have been appearing recently more and more often [38, 39, 68]. This is the time to switch from focusing mostly on the temporal dynamics. Regarding the topic of this article, first of all one need to check whether the peripheral oscillators being free of SCN control form circadian patterns of weakly synchronized oscillations. The detection of such patterns can help to establish a link between the rhythm disruption and the occurrence of cancer.

In our opinion, a new type of simulations should be able to include new experimentally discovered molecular mechanisms. Our model is designed in such a way that it can be readily modified to take account of any newly understood gene regulation processes and feedback mechanisms affecting chemo-mechanical properties of cells. In the current version of the model, we have considered only one molecular mechanism: the effect of circadian rhythms responsible for the synchronisation of cell activity in the tissue. Other mechanisms can be added to the general framework of the model. Moreover, since the present model includes individual behaviour of each cell with its own gene regulation dynamics, it can be used to simulate targeted therapeutic interventions for the various types of cancer.